\documentclass[aps,showpacs,twocolumn]{revtex4}%
\usepackage{amsfonts}
\usepackage{amsmath}
\usepackage{amssymb}
\usepackage{graphicx}%
\begin{document}
\title{Edge states and the integer quantum Hall effect of spin-chiral ferromagnetic
kagom\'{e} lattice with a general spin coupling}
\author{Zhigang Wang and Ping Zhang}
\affiliation{Institute of Applied Physics and Computational Mathematics, P.O. Box 8009,
Beijing 100088, P.R. China}

\begin{abstract}
The chiral edge states and the quantized Hall conductance (QHC) in the
two-dimensional kagom\'{e} lattice with spin anisotropies included in a
general Hund's coupling region are studied. This kagom\'{e} lattice system is
periodic in the $x$ direction but has two edges in the $y$ direction.
Numerical results show that the strength of the Hund's coupling, as well as
the spin chirality, affects the edge states and the corresponding QHC. Within
the topological edge theory, we give the expression of the QHC with the
winding number of the chiral edge states on the Riemman surface. This
expression is also compaired with that within the topological bulk theory and
they are found to keep consistent with each other.

\end{abstract}
\maketitle

\section{Introduction}

The quantum Hall effect (QHE), in which the Hall conductance (HC) $\sigma
_{xy}$ is quantized with extremely high accuracy, is a typical realization of
topological effects in condensed matter physics \cite{Klitzing}. The
topological aspect of the QHE with a periodic potential was first discussed by
Thouless, Kohmoto, Nightingale, and den Nijs \cite{TKNN}. In their work, the
HC is given by the Chern number \cite{Thouless} over the magnetic Brillouin
zone, which is a topological expression by the bulk state. Later Hatsugai
\cite{Hatsugai} suggested that the HC can be given by another topological
quantity, i.e., the \textquotedblleft winding number\textquotedblright\ of the
edge states on the complex-energy surface which is generally a high-genus
Riemann surface. To distinguish them, we call the former \textquotedblleft the
bulk theory\textquotedblright\ and the latter \textquotedblleft the edge
theory\textquotedblright. The two topological expressions for the HC, which
look quite different, actually give the same integer number.

A recent established recognition points out that the conventional QHE
originates from the non-local effect provided by the external magnetic field,
more exactly speaking, by the vector potential that describes the magnetic
field. That means a non-zero vector potential, for example, which can be
provided by spin-orbit interaction or by spin chirality, will induce the QHE
even if the external magnetic field vanishes. This type of QHE can be called
\textquotedblleft the anomalous QHE\textquotedblright%
\ \cite{Haldane,Ohgushi,Shindou}. Since Haldane performed the famous
\textquotedblleft Haldane model\textquotedblright\ in his pioneering work
\cite{Haldane} in 1988, the anomalous QHE in spin-orbit coupled
\cite{Jungwirth,Fang,Yao} or spin-chiral ferromagnetic systems
\cite{Matl,Chun,Ye,Tag} has been a hot topic in condensed matter physics.
Based on the tight-binding two-dimenisonal (2D) graphite model \cite{Semenoff}%
, Haldane model includes the next-nearest neighboring interaction and a
periodic local magnetic-flux density, which breaks time-reversal invariance
and creates a chirality. However, because the introduction of local flux is
technically difficult, it is not so easy to realize Haldane model in real
materials. Another typical spin-chiral system is the ferromagnetic system,
which is represented by pyrochlore compounds $R_{2}$Mo$_{2}$O$_{7}$ ($R$=Nd,
Sm, Gd), in which the spin configuration is noncoplanar and the spin chirality
appears. Ohgushi et al. \cite{Ohgushi} have first pointed out that the chiral
spin state can be realized by the introduction of spin anisotropy in an
\textit{ordered} spin system on the 2D kagom\'{e} lattice, which is the cross
section of the pyrochlore lattice perpendicular to the $(1,1,1)$ direction
\cite{Ramirez}. In this case, it has been shown in the topological bulk theory
\cite{Ohgushi,Wang2007} that the presence of chiral spin state may induce
gauge-invariant nonzero Chern number, thus resulting in a QHE in insulating
state. However, as stressed by Halperin \cite{Halperin}, this gauge invariance
has a relation with the edge states which are localized near the sample
boundaries. Recently we reinvestigated the kagom\'{e} lattice with boundaries.
Using the topological edge theory established by Hatsugai in the last decade
\cite{Hatsugai}, we interpreted the quantized Hall conductance (QHC) in
insulating state with the winding number of the edge states on the
complex-energy surface \cite{Edge}.

In the above-mentioned theoretical works \cite{Ohgushi, Wang2007,Edge} on the
2D kagom\'{e} lattice, a very important limit has been used. That is the
hopping electron spins are aligned with localized spins at each site of the
lattices, which is also called the \textquotedblleft infinite (\emph{strong})
Hund's coupling limit\textquotedblright. However, the latest detailed
experiments on the pyrochlores \cite{Taguchi2} showed that the spin-chiral
mechanism \emph{alone} can not explain the anomalous transport phenomena in
these systems. So in this paper we study the 2D kagom\'{e} lattice with
boundaries in a general Hund's coupling region, especially in the weak
coupling region \cite{Tatara}. We find that the edge states and QHC are not
only affected by the spin chirality, but also affected by the strength of the
Hund's coupling. Varying the spin chirality and the strength of the Hund's
coupling, two different types of phenomena are obtained. In one case that the
strength of the Hund's coupling is larger than its critical value (see
\cite{Taillefumier} or Sec. III), there are only two edge-state energies in
each bulk energy gap. With the help of the topological edge theory, we obtain
that the corresponding HC is quantized as $\sigma_{xy}$=$\pm\frac{e^{2}}{h}$
in insulating state. However, in another case that the Hund's coupling
strength is smaller than its critical value, there are \emph{possibly} four
edge-state energies in some one bulk gap. In this case, within the topological
edge theory, we obtain that the corresponding HC is quantized as $\sigma_{xy}%
$=$\pm2\frac{e^{2}}{h}$ in this insulating state, which can not occur in the
strong Hund's coupling case. We also make a comparison between these results
and those in the bulk theory \cite{Taillefumier}, and find that they keep
consistent with each other.

This paper is organized as the following. In sec. II, we introduce the
tight-binding model of the 2D kagom\'{e} lattice with boundaries along the $y$
direction and obtain the eigenvalue equations for sites. We also write out the
Hamiltonian of the 2D kagom\'{e} lattice without boundaries in the reciprocal
space in this section. Because the analytical derivation is very difficult,
then in sec. III, we numerically calculate the energy spectrum. Using the
characters of the edge-state energies in the spectrum, we study the QHE within
the topological edge theory. As comparison, we also recalculate the HC in the
infinite system in this section. Finally we make a conclusion in the last section.

\section{Model}

We consider the double-exchange ferromagnet kagom\'{e} lattice schematically
shown in Fig. 1(a). The triangle is one face of the tetrahedron. Here we
consider a pure spin model with anisotropic Dzyaloshinskii-Moriya interactions
on a kagom\'{e} lattice. It consists of an umbrella of three spins per unit
cell of the kagom\'{e} lattice. Each umbrella can be described by the
spherical coordinates of the three spins ($\pi/6,\theta$), ($5\pi/6,\theta$),
and ($-\pi/2,\theta$), as shown in Fig. 1(b). The angle $\theta$ ranges from
$0$ to $\pi$.

The tight-binding model of the 2D kagom\'{e} system can be written as the
following \cite{Taillefumier}%
\begin{equation}
H=\sum_{\langle i,j\rangle,\sigma}t_{ij}\left(  c_{i\sigma}^{\dag}c_{j\sigma
}+\text{H.c.}\right)  -J_{0}\sum_{i,\alpha,\beta}c_{i\alpha}^{\dag}\left(
\mathbf{\sigma}_{\alpha\beta}\cdot\mathbf{n}_{i}\right)  c_{i\beta},
\label{Hamiltonian}%
\end{equation}
where $t_{ij}$ is the hopping integral between two neighboring sites $i$ and
$j$; $c_{i\sigma}^{\dag}$ and $c_{i\sigma}$ are the creation and annihilation
operators of an electron with spin $\sigma$ on the site $i$. $J_{0}$ is the
effective coupling constant to each local moment $\mathbf{S}_{i}$, and these
moments are treated below as classical variables. $\mathbf{n}_{i}$ is a unit
vector collinear with the local moment $\mathbf{S}_{i}$. $\mathbf{\sigma
}_{\alpha\beta}$ are the Pauli matrices. In the following we change notation
$i\rightarrow(lms)$, where $\left(  lm\right)  $ labels the kagom\'{e} unit
cell and $s$ denote the sites A, B and C in this cell. The size of the unit
cell is set to be unit throughout this paper. This Hamiltonian has already
been discussed in the infinite Hund's coupling limit $J_{0}\rightarrow\infty$
in Refs. \cite{Ohgushi, Wang2007,Edge}. In this limit the two $\sigma
$=$\uparrow$, $\downarrow$ bands are infinitely split and the model describes
a fully polarized electron subject to a modulation of a fictitious magnetic
field, corresponding to the molecular field associated with the magnetic texture.

\begin{figure}[ptb]
\begin{center}
\includegraphics[width=0.7\linewidth]{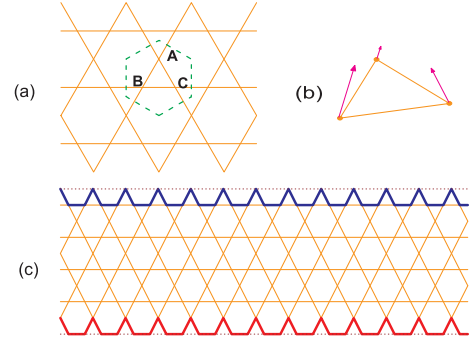}
\end{center}
\caption{(Color online) (a) 2D spin-chiral ferromagnetic kagom\'{e} lattice.
The dashed line represents the Wigner-Seitz unit cell, which contains three
independent sites (A, B, C). (b) The umbrella structure on the triangular cell
of the 2D kagom\'{e} lattice. (c) The 2D kagom\'{e} lattice system with
boundaries along the $y$ direction. The bold lines are correspond to the down
and up edges, respectively.}%
\end{figure}

Now we suppose that the system is periodic in the $x$ direction but has two
edges in the $y$ direction (see Fig. 1(c)). Since the system is periodic in
the $x$ direction, we can use a momentum representation of the electron
operator
\begin{equation}
c_{(lms\sigma)}=\frac{1}{\sqrt{L_{x}}}\sum_{k_{x}}e^{ik_{x}X_{(lms)}}%
\gamma_{ms\sigma}(k_{x}), \label{c}%
\end{equation}
where $\mathbf{R}_{(lms)}$=$\left(  X_{(lms)},Y_{(lms)}\right)  $ are the
coordinate of the site $s$ in the unit cell $(lm)$ and $k_{x}$ is the momentum
along the $x$ direction. Let us consider the \ one-particle state $|\Psi
(k_{x})\rangle$=$\sum_{ms\sigma}$ $\Psi_{ms}(k_{x})\gamma_{ms}^{\dag}%
(k_{x})|0\rangle$. Inserting it into the Schr\"{o}dinger equation
$H|\Psi\rangle$=$E|\Psi\rangle$, we can easily obtain the following three
eigenvalue equations for sites A, B, and C,%
\begin{align}
E\Psi_{mA\sigma}  &  =e^{-i\frac{k_{x}}{4}}\Psi_{mB\sigma}+e^{i\frac{k_{x}}%
{4}}\Psi_{m+1B\sigma}\nonumber\\
&  +e^{i\frac{k_{x}}{4}}\Psi_{mC\sigma}+e^{-i\frac{k_{x}}{4}}\Psi_{m+1C\sigma
}\nonumber\\
&  -\sigma J_{0}\cos\theta\Psi_{mA\sigma}+i\bar{\sigma}J_{0}\sin\theta
\Psi_{mA\bar{\sigma}},\nonumber\\
E\Psi_{mB\sigma}  &  =e^{i\frac{k_{x}}{4}}\Psi_{mA\sigma}+e^{-i\frac{k_{x}}%
{4}}\Psi_{m-1A\sigma}\nonumber\\
&  +2\cos(k_{x}/2)\Psi_{mC\sigma}\nonumber\\
&  -\sigma J_{0}\cos\theta\Psi_{mB\sigma}-J_{0}e^{i\bar{\sigma}\frac{\pi}{6}%
}\sin\theta\Psi_{mB\bar{\sigma}},\nonumber\\
E\Psi_{mC\sigma}  &  =e^{-i\frac{k_{x}}{4}}\Psi_{mA\sigma}+e^{i\frac{k_{x}}%
{4}}\Psi_{m-1A\sigma}\nonumber\\
&  +2\cos(k_{x}/2)\Psi_{mB\sigma}\nonumber\\
&  -\sigma J_{0}\cos\theta\Psi_{mC\sigma}-J_{0}e^{i\bar{\sigma}\frac{5\pi}{6}%
}\sin\theta\Psi_{mC\bar{\sigma}}. \label{Harper}%
\end{align}
Here we set $t_{ij}$=$1$ as the energy unit, $\sigma$=$\pm1$ denotes the spin
up and down, respectively. $\bar{\sigma}$=$-\sigma$. Eliminating the B- and
C-sublattice sites, one can obtain a difference equation for $\Psi_{A\sigma}$,
which is the famous Harper equation \cite{Harper}. Because the expression of
the Harper equation is too sophisticated to perform a help in analytical
resolving the edge states, we turn to make a numerical calculation and
analysis from Eq. (\ref{Harper}).

If the system is also periodic in the $y$ direction, the Hamiltonian can be
rewritten in the reciprocal space. We use the momentum representation of the
electron operator $c_{(lms\sigma)}=\frac{1}{\sqrt{L_{x}L_{y}}}\sum
_{\mathbf{k}}e^{i\mathbf{k}\cdot\mathbf{R}_{(lms)}}\gamma_{s\sigma}%
(\mathbf{k})$. Inserting the one-particle state $|\Psi(\mathbf{k})\rangle
$=$\sum_{s\sigma}$ $\Psi_{s\sigma}(\mathbf{k})\gamma_{s\sigma}^{\dag
}(\mathbf{k})|0\rangle$ into the Schr\"{o}dinger equation $H|\Psi\rangle
$=$E|\Psi\rangle$, we can easily obtain the Hamiltonian in the reciprocal
space $H(\mathbf{k})$, which is given by\begin{widetext}
\begin{equation}
H(\mathbf{k})=\left(
\begin{array}
[c]{cccccc}%
-J_{0}\cos\theta & p_{\mathbf{k}}^{1} & p_{\mathbf{k}}^{3} & iJ_{0}\sin\theta
& 0 & 0\\
p_{\mathbf{k}}^{1} & -J_{0}\cos\theta & p_{\mathbf{k}}^{2} & 0 &
-J_{0}e^{-i\frac{\pi}{6}}\sin\theta & 0\\
p_{\mathbf{k}}^{3} & p_{\mathbf{k}}^{2} & -J_{0}\cos\theta & 0 & 0 &
-J_{0}e^{-i\frac{5\pi}{6}}\sin\theta\\
-iJ_{0}\sin\theta & 0 & 0 & J_{0}\cos\theta & p_{\mathbf{k}}^{1} &
p_{\mathbf{k}}^{3}\\
0 & -J_{0}e^{i\frac{\pi}{6}}\sin\theta & 0 & p_{\mathbf{k}}^{1} & J_{0}%
\cos\theta & p_{\mathbf{k}}^{2}\\
0 & 0 & -J_{0}e^{i\frac{5\pi}{6}}\sin\theta & p_{\mathbf{k}}^{3} &
p_{\mathbf{k}}^{2} & J_{0}\cos\theta
\end{array}
\right)  ,\label{Ha}%
\end{equation}
\end{widetext}where $p_{\mathbf{k}}^{1}$=$2\cos(k_{x}/4+\sqrt{3}k_{y}/4)$,
$p_{\mathbf{k}}^{2}$=$2\cos(k_{x}/2)$ and $p_{\mathbf{k}}^{3}$=$2\cos
(-k_{x}/4+\sqrt{3}k_{y}/4)$.

\section{Edge states and the Hall conductance}

Before investigating the chiral edge states and QHC of the 2D kagom\'{e}
lattice with boundaries, we simply discuss the energy structures of this
system with different exchange interaction $J_{0}$. Firstly in the absence of
the exchange interaction, i.e., $J_{0}$=$0$, Eq. (2) becomes to be independent
of both the spin index $\sigma$ and the chiral parameter $\theta$. In this
case the two $\sigma$=$\uparrow$, $\downarrow$ bands are completely degenerate
and there are only three bulk energy bands, between which there are no bulk
energy gaps, as shown in Fig. 2. From Fig. 2, one can see that the lower
energy band becomes dispersionless ($E$=$-2$), which reflects the fact that
the 2D kagom\'{e} lattice is a line graph of the honeycomb structure
\cite{Mielke}. This flat band touches at $k_{x}$=$0$ with the middle band,
while the middle band touches at $k_{x}$=$\frac{2\pi}{3}$, $\frac{4\pi}{3}$
with the upper band. \begin{figure}[ptb]
\begin{center}
\includegraphics[width=0.6\linewidth]{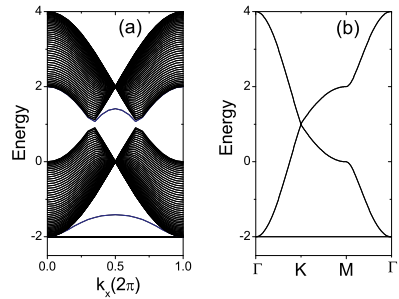}
\end{center}
\caption{(Color online) The energy spectrums of the 2D kagom\'{e} lattice (a)
with boundaries along the $y$ direction and (b) without boundaries. In both
figures, the exchange interaction $J_{0}$=$0$. There are only three energy
bands and there are \emph{no} bulk gaps between them. The lower band is flat,
which reflects that the 2D kagom\'{e} lattice is a line graph of the honeycomb
structure. In Fig. 2(a), the blue lines denote the up-edge-state energies.}%
\end{figure}

In the existence of exchange interaction, i.e., $J_{0}\neq0$, the energy
spectrum splits into two parts due to the spin-dependent potential. For very
large values of $J_{0}$, the spectrum is divided into two groups of three
bands, which is the infinite Hund's coupling case studied in Ref. \cite{Edge}.
For nonzero but not too large values of $J_{0}$, we numerically calculate the
energy spectrum and find that there are main two different phenomena for QHE
happening. The critical value of the exchange interaction $J_{c}$ depends on
the chiral parameter $\theta$. In the topological bulk theory
\cite{Taillefumier}, this critical value is analytically obtained, which reads
as
\begin{equation}
J_{c}\left(  \theta\right)  =\pm2/\sqrt{1+3\cos^{2}\theta}. \label{critical}%
\end{equation}
Although in the present model with boundaries, the critical value
$J_{c}\left(  \theta\right)  $ can not be analytically obtained, the numerical
calculations tell us that the above expression of $J_{c}\left(  \theta\right)
$ [Eq. (\ref{critical})] is also approximately valid for the present model. In
the following, we investigate the HC in both cases.

\subsection{Case I: $J_{0}<J_{c}$}

In the following we study the QHE of 2D kagom\'{e} system with boundaries in
both cases. First, we investigate the case $J_{0}<J_{c}$. As an example, we
choose the fixed chiral parameter as $\theta$=$\pi/3$, and the exchange
interaction as $J_{0}$=$1$ ($J_{c}$=$4/\sqrt{7}\approx1.51$). The number of
site A (or B, C) in the $y$ direction is chosen to be $L_{y}$=$31$. By
numerical calculating Eq. (\ref{Harper}), we draw in Fig. 3(a) the energy
spectrum in this case. From this figure, one can clearly see that the two
groups of three bulk bands are not completely divided (i.e., there is no Mott
gap) and there are two bulk gaps appearing. The range of the lower energy gap
is between $-2.68$ and $-2.6$, and that of the higher energy gap is between
$1.1$ and $1.4$, which are enlarged in Figs. 3(b) and 3(c), respectively.

In the topological edge theory \cite{Hatsugai}, when the Fermi energy lies in
one energy gap, the HC of the system is given by the winding number of the
edge states $I$, $\sigma_{xy}^{\text{edge}}$=$-\frac{e^{2}}{h}I$. The winding
number is given by the number of the intersection between the canonical loop
$\alpha_{i}$ on the Riemann surface (the complex energy surface) and the trace
of the edge-state energy $\mu_{i}$. Although in present model, the Riemann
surface is more sophisticated than that in the infinite Hund's coupling case
\cite{Edge}, we can follow the analysis in the previous work \cite{Edge} to
discuss the winding number of the edge state. Because the analytical process
is same as that in the strong Hund's coupling case, here we only present the
results as the following:

The winding number of the edge states is given by the sum of the intersection
number between the Fermi energy $\epsilon_{f}$ and the energies of the down
(or up) edge state in one period. The down (or up) edge states refer to the
eigenstates which are localized near the down (or) up boundaries of the
sample. The intersection number is obtained as: In the left neighboring of the
intersection, if the down-edge-state energy is connected with the lower energy
band, or the up-edge-state energy is connected with the upper energy band, the
intersection number is given by \textquotedblleft$+1$\textquotedblright. If
not, it is given by \textquotedblleft$-1$\textquotedblright.

Now with the above results, we investigate the HC of the system. From Fig.
3(b), one can see that when the Fermi energy $\epsilon_{f}$ lies in the lower
gap (LG), there is only one intersection between $\epsilon_{f}$ and the down
edge-state energy, which is labeled as A in Fig. 3(b). In the left neighboring
of the intersection A, the down edge-state energy is connected with the lower
band. So the winding number of the edge state is $I_{1}$=$+1$. With the same
method, one can find in Fig. 3(c) there are four edge-state energies lie in
the higher gap (HG). When the Fermi energy $\epsilon_{f}$ lies in the HG,
there are \emph{two} intersections (labeled as B and C in Fig. 3(b)) between
$\epsilon_{f}$ and the down edge-state energies. Because in the left
neighboring of the intersections, the down edge-state energies are connected
with the lower band, we can obtain that the winding number of the edge states
is $I_{2}$=$+2$. According to the topological edge theory, we can get the QHC
as
\begin{equation}
\sigma_{xy}^{\text{edge}}=\left\{
\begin{array}
[c]{c}%
-\frac{e^{2}}{h}I_{1}=-\frac{e^{2}}{h},\\
-\frac{e^{2}}{h}I_{2}=-2\frac{e^{2}}{h},
\end{array}%
\begin{array}
[c]{c}%
\epsilon_{f}\text{ }\in\text{the lower gap}\\
\epsilon_{f}\text{ }\in\text{the higher gap}%
\end{array}
\right.  . \label{xigma1}%
\end{equation}

\begin{figure}[ptb]
\begin{center}
\includegraphics[width=1.0\linewidth]{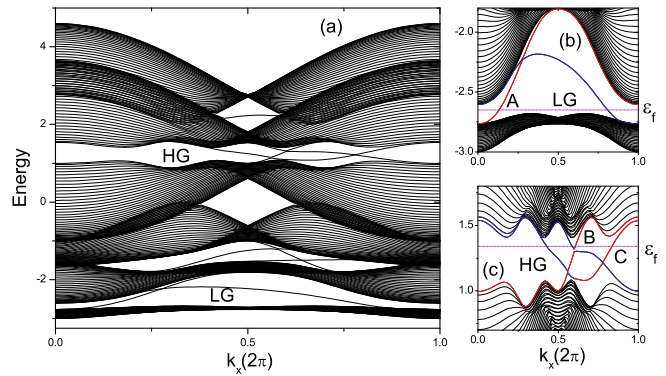}
\end{center}
\caption{(Color online) (a) The energy spectrum of 2D kagom\'{e} lattice with
edges along the $y$ direction. The parameters are chosen as $J_{0}$=$1$,
$\theta$=$\pi/3$. The shaded areas are energy bands and the lines are the
spectrum of the edge states. In this case, there is no Mott gap and there are
two energy gaps, which are enlarged in (b) and (c). In (b) and (c), the red
and blue lines correspond to the down and up edge-state energies,
respectively. }%
\end{figure}

To check the above results, we recalculate the energy spectrum and the HC of
the 2D kagom\'{e} lattice \emph{without} boundaries within the topological
bulk theory. This system has been studied by Taillefumier et al.
\cite{Taillefumier}. In the bulk theory, when the Fermi energy $\epsilon_{f}$
lies in the energy gap, the HC of the system in units of $e^{2}/h$ is given by
the sum of the Chern numbers of all the occupied energy bands \cite{Thouless}:
$\sigma_{xy}^{\text{bulk}}$=$\frac{e^{2}}{h}\sum_{n=1}^{\text{occu}}C_{n}$.
The Chern number of the $n$-th band is defined as $C_{n}$=$\frac{1}{2\pi}%
\int\Omega_{n\mathbf{k}}^{z}d^{2}\mathbf{k}$, where $\Omega_{n\mathbf{k}}^{z}%
$=$\nabla_{\mathbf{k}}\times\mathbf{A}_{n\mathbf{k}}$, and $\mathbf{A}%
_{n\mathbf{k}}$=$-i\langle u_{n\mathbf{k}}|\nabla u_{n\mathbf{k}}\rangle$ is
the geometric vector potential. With the Hamiltonian in the momentum
representation $H(\mathbf{k})$ [Eq. (\ref{Ha})], after a straightforward
numerical calculation, one can obtain the Chern numbers are $-1$, $3$, $-2$,
$-2$, $3$, $-1$ from the lowest to the topmost band. So, when the Fermi energy
lies in the LG, $\sigma_{xy}^{\text{bulk}}$=$\frac{e^{2}}{h}C_{1}$%
=$-\frac{e^{2}}{h}$; when the Fermi energy lies in the HG, $\sigma
_{xy}^{\text{bulk}}$=$\frac{e^{2}}{h}\sum_{n=1}^{4}C_{n}$=$-2\frac{e^{2}}{h}$.
Compare with Eq. (\ref{xigma1}), one obtain that $\sigma_{xy}^{\text{edge}}%
$=$\sigma_{xy}^{\text{bulk}}$. In Fig. 4(b), we plot the HC as a function of
the Fermi energy $\epsilon_{f}$ using the finite-temperature formula
$\sigma_{xy}$=$\frac{e^{2}}{h}\sum_{n}\frac{1}{2\pi}\int f_{n\mathbf{k}}%
\Omega_{n\mathbf{k}}^{z}d^{2}\mathbf{k}$, where $f_{n\mathbf{k}}$ is the Fermi
distribution function. From this figure, one can see that when the system is
in the insulating state, the HC is quantized. In the LG, the conductance is
$\sigma_{xy}$=$-\frac{e^{2}}{h}$, and in the HG, the conductance is
$\sigma_{xy}$=$-\frac{2e^{2}}{h}$. Fig. 4(a) plots the energy spectrum of the
system without boundaries.

\begin{figure}[ptb]
\begin{center}
\includegraphics[width=1.0\linewidth]{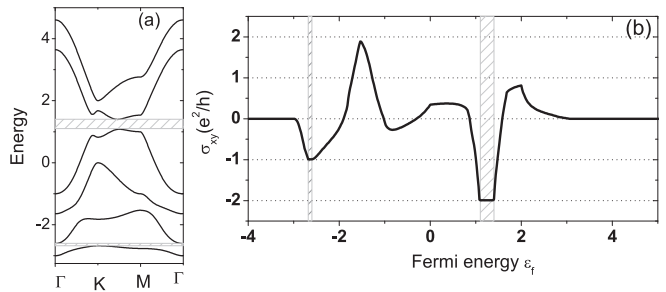}
\end{center}
\caption{(a) The energy spectrum of the 2D kagom\'{e} lattice without edges.
The parameters are same as that in Fig. 3. (b) The Hall conductance
$\sigma_{xy}$ of this system as a function of the Fermi energy $\epsilon_{f}$
at the temperature $T$=$0$. In both figures, the shaded areas are the bulk
energy gaps.}%
\label{fig1}%
\end{figure}

Note that in the strong Hund's coupling limit \cite{Edge}, the non-zero QHC
only takes two values, $\pm\frac{e^{2}}{h}$. So this is a novel result that
the HC can take the value $\pm\frac{2e^{2}}{h}$ in the weak Hund's coupling.

\subsection{Case II: $J_{0}>J_{c}$}

Then we consider the other case, in which $J_{0}>J_{c}$. As an example, we
choose the fixed chiral parameter $\theta$=$\pi/3$ and the exchange
interaction $J_{0}$=$2$. The number of sites A (or B, C) in the $y$ direction
is also chosen to be $L_{y}$=$31$. With Eq. (\ref{Harper}), we plot in Fig.
5(a) the energy spectrum of the 2D kagom\'{e} lattice with boundaries. From
Fig. 5(a), one can find that there are four gaps in this case. We call them
G-I, G-II, G-III (which is the Mott gap), and G-IV corresponding to that
between $-3.54$ and $-3.36$, between $-2.0$ and $-1.65$, between $-1$ and $0$,
and between $1.65$ and $2.0$, respectively. To clearly see these gaps and the
edge states, we enlarge the G-I, -II, and -IV in Figs. 5(b)-(d), respectively.

From Figs. 5(b)-(d), one can see that when the Fermi energy $\epsilon_{f}$
lies in the G-I (or -II, -IV), there is only one intersection between
$\epsilon_{f}$ and the down edge-state energy, which is labeled as point A (B,
C) in Fig. 5(b) (5(c), 5(d)). In the left neighboring of the intersection A
(B), the down edge-state energy is connected with the lower band. So the
winding number of the edge state is $I_{\text{I(II)}}$=$+1$. But in the left
neighboring of the intersection C, the down edge-state energy is connected
with the upper band. The winding number of the edge state is $I_{\text{IV}}%
$=$-1$. By the way, since there is no edge-state energy lies in the Mott gap,
there is no intersection between $\epsilon_{f}$ and the left edge-state energy
when $\epsilon_{f}$ lies in the Mott gap, and correspondingly, the winding
number is $I_{\text{III}}$=$0$. So, with the topological edge theory, one can
easily obtain that when the Fermi energy $\epsilon_{f}$ lies in the bulk gaps,
the QHC is%
\begin{equation}
\sigma_{xy}^{\text{edge}}=\left\{
\begin{array}
[c]{c}%
-\frac{e^{2}}{h}I_{\text{I}}=-\frac{e^{2}}{h},\\
-\frac{e^{2}}{h}I_{\text{II}}=-\frac{e^{2}}{h},\\
-\frac{e^{2}}{h}I_{\text{III}}=0,\\
-\frac{e^{2}}{h}I_{\text{IV}}=\frac{e^{2}}{h},
\end{array}%
\begin{array}
[c]{c}%
\epsilon_{f}\text{ }\in\text{G-I}\\
\epsilon_{f}\text{ }\in\text{G-II}\\
\epsilon_{f}\text{ }\in\text{the Mott gap}\\
\epsilon_{f}\text{ }\in\text{G -IV}%
\end{array}
\right.  .\label{xigma2}%
\end{equation}
\begin{figure}[ptb]
\begin{center}
\includegraphics[width=1.0\linewidth]{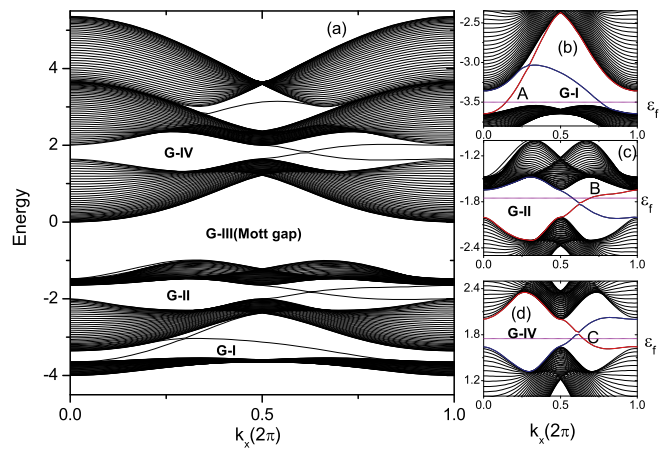}
\end{center}
\caption{(Color online) (a) The energy spectrum of 2D kagom\'{e} lattice with
edges along the $y$ direction. The parameters are chosen as $J_{0}$=$2$,
$\theta$=$\pi/3$. The shaded areas are energy bands and the lines are the
spectrum of the edge states. In this case, there is a Mott gap. Besides that,
there are three other energy gaps, which are enlarged in (b), (c) and (d). In
(b)-(d), the red and blue lines correspond to the down and up edge-state
energies, respectively. }%
\end{figure}\begin{figure}[ptbptb]
\begin{center}
\includegraphics[width=1.0\linewidth]{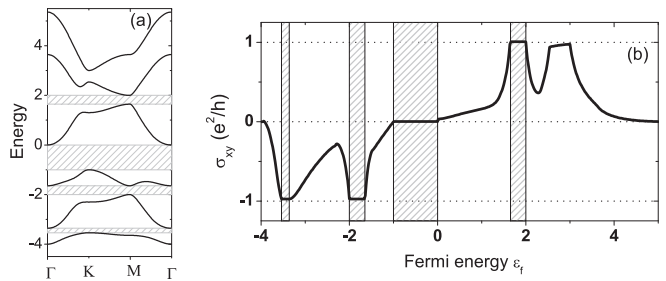}
\end{center}
\caption{(a) The energy spectrum of the 2D kagom\'{e} lattice without edges.
The parameters are same as that in Fig. 5. (b) The Hall conductance
$\sigma_{xy}$ of this system as a function of the Fermi energy $\epsilon_{f}$
at the temperature $T$=$0$. \ In both figures, the shaded areas are the bulk
energy gaps.}%
\end{figure}

Similar to the discussion in case-I, to check the above results, we also
recalculate the HC of the corresponding 2D kagom\'{e} lattice \emph{without}
boundaries. After a straightforward numerical calculation, one obtains that
the Chern numbers are $-1$, $0$, $1$, $1$, $0$, $-1$ from the lowest to the
topmost band. When the Fermi energy $\epsilon_{f}$ lies in G-I, only the
lowest band is fully filled. So the HC is $\sigma_{xy}^{\text{I},\text{bulk}}%
$=$\frac{e^{2}}{h}C_{1}$=$-\frac{e^{2}}{h}$; When $\epsilon_{f}$ lies in G-II,
the lowest two bands are fully occupied, $\sigma_{xy}^{\text{II},\text{bulk}}%
$=$\frac{e^{2}}{h}\left(  C_{1}+C_{2}\right)  $=$-\frac{e^{2}}{h}$; When
$\epsilon_{f}$ lies in the Mott gap, $\sigma_{xy}^{\text{III},\text{bulk}}%
$=$\frac{e^{2}}{h}\sum_{n=1}^{3}C_{n}$=$0$; When $\epsilon_{f}$ lies in G-IV,
the lowest four bands are fully occupied, $\sigma_{xy}^{\text{IV},\text{bulk}%
}$=$\frac{e^{2}}{h}\sum_{n=1}^{4}C_{n}$=$\frac{e^{2}}{h}$. Comparing with Eq.
(\ref{xigma2}), one can also obtain that $\sigma_{xy}^{\text{edge}}$%
=$\sigma_{xy}^{\text{bulk}}$. In Fig. 6(b) we plot the HC as a function of the
Fermi energy $\epsilon_{f}$. From this figure, one can see $\sigma_{xy}$ is
similar to that in the strong Hund's coupling limit. The strength $J_{0}$ is
more larger, the results more tend to those in the strong Hund's coupling
limit. Fig. 6(a) plots the energy spectrum of the system without boundaries.

\section{Summary}

In summary, in this paper we have studied the chiral edge states and the QHC
of the 2D kagom\'{e} lattice with spin anisotropies included in a general
Hund's coupling region. This system is periodic in the $x$ direction but has
two edges in the $y$ direction. By numerical calculation, we find that both
the strength of the Hund's coupling and the spin chirality affect the edges
states and the corresponding QHC. Upon varying the chirality and the strength
of the Hund's coupling, two types of phenomena occur, which are distinguished
by the critical relation between these two parameters (Eq. (\ref{critical})).
A remarkable difference with the infinite Hund's coupling limit is that there
are possibly four edge-state energies in one bulk gap in the case $J_{0}%
<J_{c}$. If the Fermi energy lies in this gap, the HC is quantized as
$\sigma_{xy}$=$\pm2\frac{e^{2}}{h}$. We also give the corresponding expression
of the QHC without boundaries within the topological bulk theory. Both two
topological expressions give the same QHC.

\begin{acknowledgments}
This work was supported by NSFC under Grants Nos. 10604010 and 60776063.
\end{acknowledgments}

\end{document}